\documentclass[aps,twocolumn,a4paper,nofootinbib,superscriptaddress]{revtex4}
\usepackage[dvips]{graphicx}
\usepackage{amsmath}
\usepackage{amssymb}

\begin{document}

\title{Many-body effects of Coulomb interaction on Landau levels in graphene}
\author{A. A. Sokolik}%
\affiliation{Institute for Spectroscopy, Russian Academy of Sciences, 142190 Troitsk, Moscow, Russia}%
\affiliation{National Research University Higher School of Economics, 109028 Moscow, Russia}%
\author{A. D. Zabolotskiy}%
\affiliation{Dukhov Research Institute of Automatics (VNIIA), 127055 Moscow, Russia}%
\author{Yu.\ E. Lozovik}%
\email{lozovik@isan.troitsk.ru}
\affiliation{Institute for Spectroscopy, Russian Academy of Sciences, 142190 Troitsk, Moscow, Russia}%
\affiliation{National Research University Higher School of Economics, 109028 Moscow, Russia}
\affiliation{Dukhov Research Institute of Automatics (VNIIA), 127055 Moscow, Russia}%

\begin{abstract}
In strong magnetic fields, massless electrons in graphene populate relativistic Landau levels with the square-root
dependence of each level energy on its number and magnetic field. Interaction-induced deviations from this
single-particle picture were observed in recent experiments on cyclotron resonance and magneto-Raman scattering.
Previous attempts to calculate such deviations theoretically using the unscreened Coulomb interaction resulted in
overestimated many-body effects. This work presents many-body calculations of cyclotron and magneto-Raman transitions
in single-layer graphene in the presence of Coulomb interaction, which is statically screened in the random-phase
approximation. We take into account self-energy and excitonic effects as well as Landau level mixing, and achieve good
agreement of our results with the experimental data for graphene on different substrates. Important role of a
self-consistent treatment of the screening is found.
\end{abstract}

\maketitle

\section{Introduction}

Graphene, the monolayer two-dimensional carbon crystal, grants a possibility to study how a many-body system of
massless Dirac electrons behave in electric and magnetic fields
\cite{CastroNeto,Goerbig,Novoselov,Orlita,Basov,Miransky}. One of the most striking manifestations of the
``relativistic'' nature of graphene is the unconventional half-integer quantum Hall effect in strong magnetic field
\cite{Novoselov}. The role of Coulomb interaction in graphene in the quantum Hall regime is still a debatable and
controversial topic \cite{Goerbig,Iyengar,Bychkov,Barlas,Roldan1,Shizuya1,Shizuya2,Gorbar,Menezes,Miller,Luican,Chae,
Yu,Jiang1,Henriksen,Jiang2,Chen,Takehana,Faugeras,Berciaud,Orlita,Basov,Miransky,Shovkovy}.

Landau levels of electrons in graphene have the non-equidistant energies \cite{Goerbig}
\begin{eqnarray}
E_n^{(0)}=\mathrm{sgn}(n)\frac{v_\mathrm{F}}{l_H}\sqrt{2|n|},\quad n=0,\pm1,\pm2,\ldots,\label{E_0}
\end{eqnarray}
where $v_\mathrm{F}\approx10^6\,\mbox{m/s}$ is the Fermi velocity, $l_H=\sqrt{c/|e|B}$ is the magnetic length
(hereafter we set $\hbar\equiv1$). Contrary to the case of a non-relativistic electron gas, the energies of cyclotron
transitions between Landau levels in graphene are not protected by Kohn's theorem \cite{Kohn} against interaction
induced corrections, as both predicted theoretically
\cite{Iyengar,Bychkov,Barlas,Roldan1,Shizuya1,Shizuya2,Gorbar,Menezes,Shovkovy} and reported in experimental works
\cite{Jiang1,Henriksen,Jiang2,Chen,Takehana,Luican,Chae,Yu,Berciaud,Faugeras}.

The following major signatures of Coulomb many-body effects are observed: a) the energy of $0\rightarrow1$ or
$-1\rightarrow0$ (referred to as $\mathrm{T}_1$) cyclotron inter-Landau level transitions and that of $-1\rightarrow2$
or $-2\rightarrow1$ transitions ($\mathrm{T}_2$) have the ratio deviating from the single-particle prediction
$1/(1+\sqrt2)$ \cite{Jiang1,Henriksen,Jiang2}; b) the renormalized Fermi velocities $v_\mathrm{F}^*$ which characterize
the energies of symmetric interband $-n\rightarrow n$ transitions (referred to as $\mathrm{L}_n$) measured in
magneto-Raman scattering \cite{Berciaud,Faugeras} demonstrate significant dependence on magnetic field and on a
substrate dielectric constant.

The existing theoretical calculations \cite{Iyengar,Bychkov,Barlas,Roldan1,Shizuya1,Shizuya2,Menezes,Faugeras} of
inter-Landau level transitions in graphene were carried out in the first order in Coulomb interaction, which implies
using of the unscreened Coulomb interaction in all matrix elements \cite{Note1}. This results in the overestimation of
many-body effects in comparison with the experimental data, as noted in Refs.
\cite{Shizuya1,Shizuya2,Jiang1,Jiang2,Faugeras,Chae}.

In this article, we calculate the energies of the cyclotron and magneto-Raman transitions between Landau levels in
graphene using the Coulomb interaction which is screened in the random-phase approximation in the static limit. We
include exchange self-energy and excitonic contributions to the transition energies, as well as the Landau level mixing
in the excitonic channel, and fit the experimental data \cite{Jiang1,Jiang2,Faugeras,Berciaud} with our calculations.
Using the bare Fermi velocity $v_\mathrm{F}=0.85\times10^6\,\mbox{m/s}$ and realistic dielectric constants, we have
achieved much better agreement with both the magneto-Raman \cite{Berciaud,Faugeras} and cyclotron resonance
\cite{Jiang1,Jiang2} experiments than in previous attempts of other authors, which had dealt with the unscreened
interaction \cite{Shizuya1,Shizuya2,Jiang1,Faugeras}. Moreover, we find the important role of a self-consistent
suppression of the screening due to an upward renormalization of transition energies.

The article is organized as follows. Our theoretical model is introduced in Sec.~\ref{sec2}. The results of
calculations are presented and the experimental data are fitted in Sec.~\ref{sec3}. Finally, the conclusions are made
in Sec.~\ref{sec4}.

\section{Theoretical model}
\label{sec2}

\subsection{Exchange self-energies}

Most theoretical models of Coulomb many-body effects in graphene
\cite{Iyengar,Bychkov,Barlas,Roldan1,Shizuya1,Shizuya2,Gorbar,Menezes,Faugeras,Shovkovy} take into account three major
contributions to the inter-Landau level transition energies: single-particle exchange self-energies of electron and
hole, an excitonic shift due to electron-hole Coulomb attraction (also referred to as a vertex correction) and an
electron-hole exchange energy. The latter contribution is principal in calculating dispersions of collective
magnetoplasmon excitations \cite{Iyengar,Bychkov,Roldan1,Shizuya3,Berman,Tahir,Roldan2,Lozovik,Roldan3,Goerbig}, but
vanishes for optically excited nearly zero-momentum electron-hole pairs, therefore we will not include it in our
calculations.

Renormalization of single-particle energy levels is conventionally treated using the Hartree-Fock approximation and the
unscreened Coulomb interaction. Omitting the Hartree term, which affects only electrostatics of a graphene layer, we
get the single-particle energy levels
\begin{eqnarray}
E_n^\mathrm{(HF)}=E_n^{(0)}+\Sigma_n^\mathrm{(exch)}\label{E_HF}
\end{eqnarray}
shifted due to the Fock exchange self-energies (see Refs.~\cite{Iyengar,Bychkov,Roldan1,Faugeras} for the details of
calculations)
\begin{eqnarray}
\Sigma_n^\mathrm{(exch)}=-\sum_{n'k'}f_{n'}\langle\psi_{nk},\psi_{n'k'}|v|\psi_{nk},\psi_{n'k'}\rangle.\label{Sigma_exch}
\end{eqnarray}
Here $f_{n'}$ is the occupation number of the $n'$-th Landau level ($0\leqslant f_{n'}\leqslant1$),
$\langle\psi_{nk},\psi_{n'k'}|v|\psi_{nk},\psi_{n'k'}\rangle$ is the exchange matrix element of the Coulomb interaction
$v$; the latter has the form $v(\mathbf{r})=e^2/\varepsilon|\mathbf{r}|$ in a surrounding medium with the dielectric
constant $\varepsilon$. Each single-particle state $\psi_{nk}$ is specified by a Landau level number $n$ and a guiding
center index $k=0,1,2,\ldots$, when we work in the symmetric gauge.

As known \cite{Iyengar,Bychkov,Barlas,Roldan1,Shizuya1,Shizuya2,Gorbar}, the self-energies (\ref{Sigma_exch}) diverge
logarithmically when calculating the sum over the filled Landau levels of negative energies, so the cutoff
$n'\geqslant-n_\mathrm{c}$ is required in order to obtain finite results. The value of $n_\mathrm{c}$ can be estimated
by equating the concentration $gn_\mathrm{c}/2\pi l_H^2$ of electrons on $n_\mathrm{c}$ Landau levels (with taking into
account the fourfold spin and valley degeneracy $g=4$) to that in the filled valence band of intrinsic graphene,
$2/S_0$:
\begin{eqnarray}
n_\mathrm{c}=\frac{\pi l_H^2}{S_0}\approx\frac{39600}{B\,\mbox{[T]}}.
\end{eqnarray}
Here $S_0=a^2\sqrt3/2$ is the area of graphene elementary cell, $a\approx2.46\,\mbox{\AA}$. Separating the part of
(\ref{Sigma_exch}) which diverges in the $n_\mathrm{c}\rightarrow\infty$ limit, we get
\begin{equation}
\Sigma_n^\mathrm{(exch)}=\frac{e^2}{\varepsilon l_H}\left\{-\sqrt{\frac{n_\mathrm{c}}2}+\mathrm{sgn}(n)
\frac{\sqrt{|n|}}{4\sqrt2}\ln\frac{n_\mathrm{c}}{|n|}\right\}+\mathcal{O}(1).\label{Sigma_exch2}
\end{equation}

In the weak magnetic field limit $B\rightarrow0$, (\ref{Sigma_exch2}) can be tracked to the well-known form of electron
self-energy in graphene in the absence of magnetic field. The first term of (\ref{Sigma_exch2}) equals to the large
negative constant part of the Hartree-Fock self-energy $-e^2p_\mathrm{c}/2\varepsilon$ \cite{Roldan1,Sokolik,Hwang},
where $p_\mathrm{c}=\sqrt{2\pi/S_0}$ is the cutoff momentum. The second term describes the logarithmic renormalization
of the Fermi velocity \cite{Gonzalez,CastroNeto}
\begin{eqnarray}
v_\mathrm{F}^*=v_\mathrm{F}+\frac{e^2}{4\varepsilon}\ln\frac{p_c}{p_\mathrm{F}},
\end{eqnarray}
where $p_\mathrm{F}$ is the Fermi momentum. Thus the exchange self-energies with and without magnetic field have the
same cutoff dependencies up to the linear and logarithmic levels. The same result was obtained in Ref.~\cite{Menezes} by a
different method.

\subsection{Excitonic effects}

An excitonic energy shift due to Coulomb attraction between electron on the $n_1$-th Landau level and hole on the
$n_2$-th level is often calculated in the first order in Coulomb interaction
\cite{Iyengar,Shizuya1,Shizuya2,Faugeras,Roldan1}:
\begin{eqnarray}
\Delta E_{n_1n_2}^\mathrm{(exc)}=-\langle\Phi_{n_1n_2}|v|\Phi_{n_1n_2}\rangle,\label{E_exc}
\end{eqnarray}
where $|\Phi_{n_1n_2}\rangle$ is the noninteracting electron-hole (or magnetoexcitonic) state at zero momentum.

Eq.~(\ref{E_exc}) is the simplification of the more general picture, where the mixing of different $n_2\rightarrow n_1$
transitions should occur in the excitonic ladder \cite{Iyengar,Bychkov,Lozovik}. To take it into account, one must
consider the Hamiltonian in the basis of noninteracting electron-hole states with the matrix elements given by
\begin{eqnarray}
\langle\Phi_{n_1'n_2'}|H|\Phi_{n_1n_2}\rangle = \delta_{n_1n_1'}\delta_{n_2n_2'}(E_{n_1}-E_{n_2})\nonumber\\
-\langle\Phi_{n_1'n_2'}|v|\Phi_{n_1n_2}\rangle,\label{mixing}
\end{eqnarray}
and find its eigenvalues. Here $E_n$ are the single-particle energies which are already renormalized by Coulomb
interaction.

Our estimates show that it is sufficient to consider only the mixing of $\mathrm{L}_1$ and $\mathrm{L}_2$ magneto-Raman
transitions or $\mathrm{T}_1$ and $\mathrm{T}_2$ cyclotron transitions, and its major effect is a slight increase of
the energy distance between these transition lines due to an interlevel repulsion. The mixing with higher lying
transitions has an overall weak effect in the presence of the screening (although it was estimated to be significant in
the absence of the screening at $\varepsilon\sim1$ \cite{Lozovik}).

\begin{figure}[t]
\begin{center}
\resizebox{0.85\columnwidth}{!}{\includegraphics{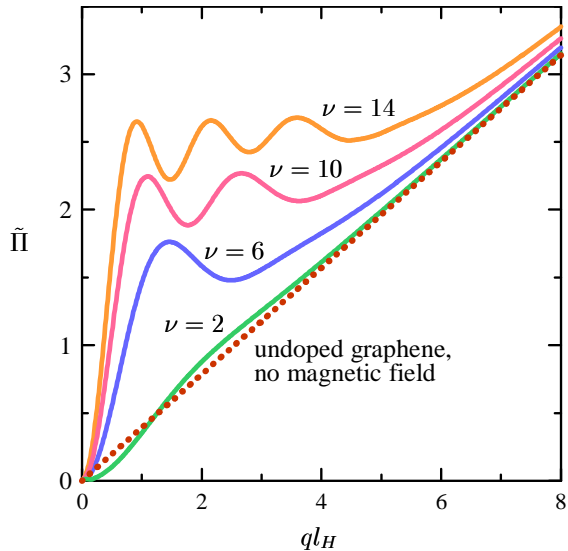}}
\end{center}
\caption{\label{Fig1}Dimensionless static polarizability of graphene in magnetic field $\tilde\Pi$ calculated in the
random-phase approximation at zero temperature as functions of the dimensionless momentum $ql_H$. The filling factors
$\nu=2,6,10,14$ correspond to complete fillings of, respectively, $n=0,1,2,3$ Landau levels. For comparison,
polarizability of undoped graphene without magnetic field $\tilde\Pi=\pi ql_H/8$ is shown by the dotted line.}
\end{figure}

\subsection{Screening of Coulomb interaction}

Polarization of the electron gas in graphene, which can be described in terms of virtual electron-hole pairs, leads to
the screening of Coulomb interaction. The screened interaction in the static limit $\omega\rightarrow0$ is
\begin{eqnarray}
V(q)=\frac{v(q)}{1-v(q)\Pi(q,0)},\label{V_scr}
\end{eqnarray}
where $v(q)=2\pi e^2/\varepsilon q$ is the two-dimensional Fourier transform of the unscreened interaction, and
$\Pi(q,\omega)$ is the irreducible polarizability. In the random-phase approximation,
\begin{eqnarray}
\Pi(q,0)=g\sum_{nn'}F_{nn'}(q)\frac{f_n-f_{n'}}{E_n^{(0)}-E_{n'}^{(0)}},\label{Pol}
\end{eqnarray}
where $F_{nn'}(q)$ are the Landau level form factors (see the details of polarizability calculations in Refs.
\cite{Goerbig,Gumbs,Lozovik,Pyatkovskiy,Roldan2,Roldan3}).

Introducing the positively valued dimensionless polarizability $\tilde\Pi(ql_H)=-(2\pi v_\mathrm{F}l_H/g)\Pi(q,0)$, we
get Eq.~(\ref{V_scr}) in the form
\begin{eqnarray}
V(q)=\frac{v(q)}{1+gr_\mathrm{s}\tilde\Pi(ql_H)/ql_H},\label{V_scr1}
\end{eqnarray}
where
\begin{eqnarray}
r_\mathrm{s}=\frac{e^2}{\varepsilon v_\mathrm{F}}.\label{rs}
\end{eqnarray}
The dimensionless parameter $r_\mathrm{s}$ is conventionally used to characterize a relative strength of Coulomb
interaction, but in our approach it appears only as a multiplier of $\tilde\Pi$ in the denominator of (\ref{V_scr1})
and thus characterizes the relative strength of the screening. The case of unscreened Coulomb interaction corresponds
to the zeroth order in this parameter, $r_\mathrm{s}=0$.

The numerically calculated polarizabilities are shown in Fig.~\ref{Fig1} at different integer fillings $\nu$ of Landau
levels ($\nu=0$ in undoped graphene and $\nu=4n+2$ when the $n$-th highest occupied level is completely filled). The
functions $\tilde\Pi(ql_H)$ oscillate at $ql_H\sim1$ reflecting the nodal structure of Landau level wave functions and
tend to the polarizability \cite{CastroNeto} of undoped graphene without magnetic field $\tilde\Pi=\pi ql_H/8$ at
$ql_H\gg1$ (see also analysis in \cite{Gumbs,Roldan3}). Due to the electron-hole symmetry of the Dirac model,
$\tilde\Pi$ does not depend on the sign of $\nu$. Remarkably, the same symmetry makes the polarizability independent on
the filling of the 0-th Landau level, thus $\tilde\Pi$ is the same for $\nu=0$ and $\nu=\pm2$, if we neglect intralevel
electron transitions.

To improve calculations of the transition energies with taking into account the screening, we replace $v$ with
(\ref{V_scr1}) in the matrix elements of Coulomb interaction when calculate the electron self-energies
(\ref{Sigma_exch}) and treat excitonic effects (\ref{mixing}):
\begin{eqnarray}
\langle\Phi_{n_1'n_2'}|H|\Phi_{n_1n_2}\rangle = \delta_{n_1n_1'}\delta_{n_2n_2'}(E_{n_1}-E_{n_2})\nonumber\\
-\langle\Phi_{n_1'n_2'}|V|\Phi_{n_1n_2}\rangle,\label{mixing_Scr}\\
E_n=E_n^{(0)}-\sum_{\substack{n'k'\\ n'\geqslant
-n_\mathrm{c}}}f_{n'}\langle\psi_{nk},\psi_{n'k'}|V|\psi_{nk},\psi_{n'k'}\rangle.\label{Sigma_Scr}
\end{eqnarray}
Eqs.~(\ref{Pol})--(\ref{Sigma_Scr}) compose the starting point of our numerical calculations.

\section{Calculation results and comparison with experiments}
\label{sec3}

\subsection{Magneto-Raman transitions}

\begin{figure}[t]
\begin{center}
\resizebox{0.9\columnwidth}{!}{\includegraphics{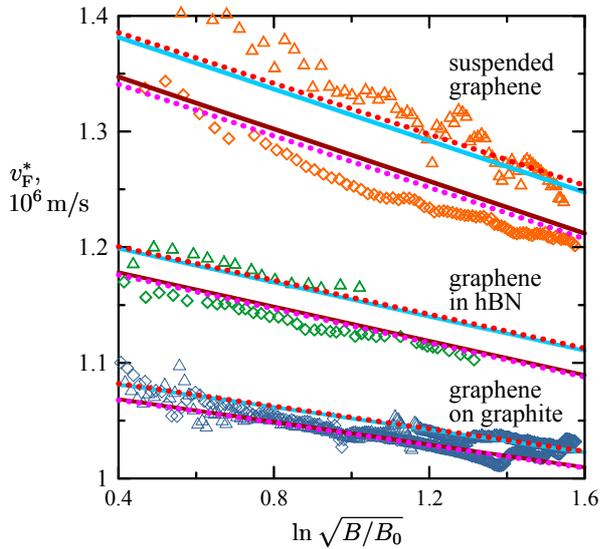}}
\end{center}
\caption{\label{Fig2}Renormalized Fermi velocities $v_\mathrm{F}^*$ as functions of magnetic field $B$ (the reference
field strength is $B_0=1\,\mbox{T}$). Diamonds and triangles: experimental points from Ref.~\cite{Faugeras} for
$\mathrm{L}_1$ and $\mathrm{L}_2$ transitions in three graphene samples. Solid and dotted lines: calculations with,
respectively, unscreened and screened Coulomb interactions (see their parameters in Table~\ref{Table}) for the same
pair of transitions in each sample.}
\end{figure}

The recent experiments \cite{Berciaud,Faugeras} on magneto-Raman scattering showing clear signs of Coulomb many-body
effects were carried out with undoped ($\nu=0$) graphene on three types of substrate: 1) suspended graphene, 2)
graphene encapsulated in hexagonal boron nitride (hBN), 3) graphene on graphite. In each sample, the energies
$E_{-1\rightarrow 1}$ and $E_{-2\rightarrow 2}$ of two transitions $\mathrm{L}_1$ and $\mathrm{L}_2$ were measured as
functions of magnetic field in the range $2\div25\,\mbox{T}$ and then converted to the renormalized Fermi velocities
$v_\mathrm{F}^*$. The latter describe fictitious single-particle Landau levels (\ref{E_0}) having the same energy
difference: $2\sqrt{2n}v_\mathrm{F}^*/l_H=E_{-n\rightarrow n}$. The experimental points $v_\mathrm{F}^*(B)$ are
reproduced in Fig.~\ref{Fig2}.

Our many-body calculations of $v_\mathrm{F}^*$ require the bare Fermi velocity $v_\mathrm{F}$ and the dielectric
constant $\varepsilon$ as input parameters. Otherwise, these quantities can be obtained by least square fitting of
experimental points. Generally, the best approximations to the experimental data can be achieved by independent
adjustment of $v_\mathrm{F}$ and $\varepsilon$ for each of six transition lines in Fig.~\ref{Fig2}. We use more
realistic fitting procedure, adjusting separate values of $\varepsilon$ for the pair of transitions $\mathrm{L}_{1,2}$
in each graphene sample as well as a common $v_\mathrm{F}$ for all samples.

Our attempts to fit the experimental data in different approximations (see below) show that the optimal value of common
$v_\mathrm{F}$ is about $0.85\times10^6\,\mbox{m/s}$. Smaller or larger values of $v_\mathrm{F}$ do not allow us to
reproduce the slopes of experimental dependencies of $v_\mathrm{F}^*$ on $\ln\sqrt{B/B_0}$ accurately enough for all
samples simultaneously. Assuming this $v_\mathrm{F}$, then we adjust the effective dielectric constant $\varepsilon$
for each graphene sample in order to achieve the best least square fit for its pair $\mathrm{L}_{1,2}$ of transitions.
In agreement with the experiment, our calculations show that $v_\mathrm{F}^*$ for $\mathrm{L}_2$ is always higher than
for $\mathrm{L}_1$ in the same sample, and dependencies of $v_\mathrm{F}^*$ on $\ln\sqrt{B/B_0}$ are approximately
linear.

\begin{figure}[t]
\begin{center}
\resizebox{0.9\columnwidth}{!}{\includegraphics{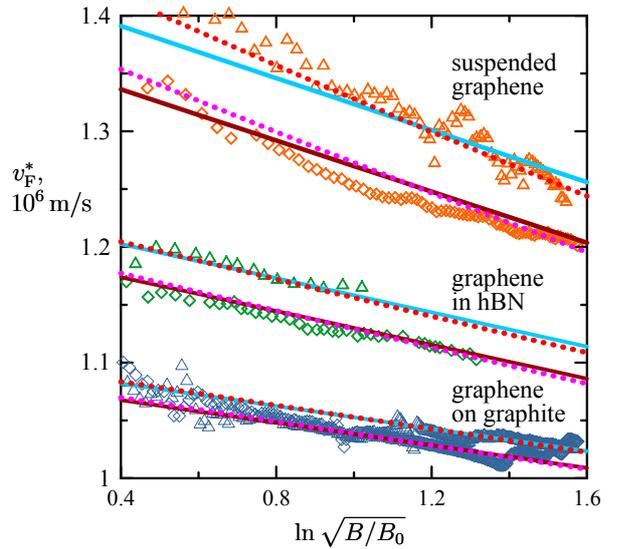}}
\end{center}
\caption{\label{Fig3}The same as Fig.~\ref{Fig2}, but solid and dotted lines present calculations with a
``self-consistent'' screening using, respectively, constant and varying $r_{\mathrm{s},n}$ (see the calculation
parameters in Table~\ref{Table}).}
\end{figure}

First, we make the adjustment of $\varepsilon$ using the unscreened interaction, with the results shown in
Fig.~\ref{Fig2} (solid lines) and Table~\ref{Table} (first column). The obtained $\varepsilon$ turned out to be too
large in comparison with the actual dielectric constants of suspended graphene ($\varepsilon\approx1$) and graphene in
hBN ($\varepsilon\approx4.5$) because they need to mimic the screening caused by both surrounding medium and graphene
electrons. Besides, the distance between the $\mathrm{L}_1$ and $\mathrm{L}_2$ lines is clearly insufficient in this
approximation. The same conclusions was made in Ref.~\cite{Faugeras}, where $\varepsilon$ was adjusted to approximate
the slopes of experimental dependencies of $v_\mathrm{F}^*$ on $\ln\sqrt{B/B_0}$ and additional fictitious
$\varepsilon_{\delta v}$ was needed to reproduce the interline distance in each sample.

Then we use the statically screened interaction and obtain the fitting results shown in Fig.~\ref{Fig2} (dotted lines)
and Table~\ref{Table} (second column). Although we still have insufficient interline distances, the resulting
$\varepsilon$ are no longer overestimated with respect to the actual ones, but are even underestimated (for example,
$\varepsilon$ is even unphysically smaller than 1 in the case of suspended graphene). A possible reason is
overestimation of the screening in the static limit in comparison with a full dynamical screening.

In order to improve agreement between theory and experiment, we can make the screening approximately
``self-consistent''. Indeed, if the renormalized Fermi velocity
$v_\mathrm{F}^*\approx(1.05\div1.4)\times10^6\,\mbox{m/s}$, which describes observable energies of inter-Landau level
transitions, is 25\%--65\% higher than the bare velocity $v_\mathrm{F}=0.85\times10^6\,\mbox{m/s}$, then the
polarizability $\Pi(q,0)$ should correspondingly be renormalized to lower values due to increased energy denominators
in (\ref{Pol}). Since a full self-consistent treatment of renormalized transition energies in the polarizability would
be computationally demanding and beyond the accuracy level of the static random-phase approximation, we resort to a
simplified semi-phenomenological approach. We still use the same dimensionless $\tilde\Pi(ql_H)$ (see Fig.~\ref{Fig1}),
which describes qualitative features of the screening, but change the value of the parameter $r_\mathrm{s}$, which
determines quantitatively the overall screening strength. For each transition $\mathrm{L}_n$ in each sample, we
substitute an averaged (over the magnetic field range) value of experimental $v_\mathrm{F}^*$ instead of the bare
$v_\mathrm{F}$ into the formula (\ref{rs}): $r_{\mathrm{s},n}=e^2/\varepsilon\langle v_{\mathrm{F},n}^*\rangle$. This
replacement reduces the resulting $r_{\mathrm{s},n}$ and effectively weakens the screening.

The results of this approach are shown in Fig.~\ref{Fig3} (solid lines) and Table~\ref{Table} (third column). Increased
interline distances provide much better agreement with the experimental data, that indicates importance of a
self-consistent treatment of the screening.

We can improve agreement with the experiment even further, if take into account significant change of
$v_{\mathrm{F},n}^*$ in the experimental range of magnetic field, which is most noticeable in the case of suspended
graphene. In this fourth approximation, we assume $r_{\mathrm{s},n}=e^2/\varepsilon v_{\mathrm{F},n}^*(B)$, where
$v_{\mathrm{F},n}^*(B)$ is the linear fit to experimental data for the $\mathrm{L}_n$ transition. The results, shown in
Fig.~\ref{Fig3} (dotted lines) and Table~\ref{Table} (fourth column), demonstrate the best agreement with the
experiment both in line slopes and interline distances.

\begin{table*}[t]
\centering
\begin{tabular}{|l||c|c|c|c|}
\hline%
&Unscreened&Screened&Self-consistent&Self-consistent\\
Sample&interaction&interaction&screening&screening\\
&$r_{\mathrm{s},n}=0$&$r_{\mathrm{s},n}=e^2/\varepsilon v_\mathrm{F}$&$r_{\mathrm{s},n}=e^2/\varepsilon\langle
v_{\mathrm{F},n}^*\rangle$&
$r_{\mathrm{s},n}=e^2/\varepsilon v_{\mathrm{F},n}^*(B)$\\
\hline%
Suspended graphene&\hphantom{1}4.88&0.90&2.20&2.22\\
\hline
Graphene in hBN&\hphantom{1}7.41&3.41&4.47&4.48\\
\hline
Graphene on graphite&11.16&7.14&7.87&7.88\\
\hline
\end{tabular}
\caption{\label{Table}Dielectric constants of surrounding media, obtained by least square fittings of magneto-Raman
experimental data from Ref.~\cite{Faugeras} for each graphene sample at $v_\mathrm{F}=0.85\times10^6\,\mbox{m/s}$. The
fittings are carried out in four different theoretical approximations (see the text), and the corresponding screening
parameters $r_{\mathrm{s},n}$ for each transition $\mathrm{L}_n$, which are used in these approximations, are shown.}
\end{table*}

\subsection{Cyclotron transitions}

\begin{figure}[b]
\begin{center}
\resizebox{0.9\columnwidth}{!}{\includegraphics{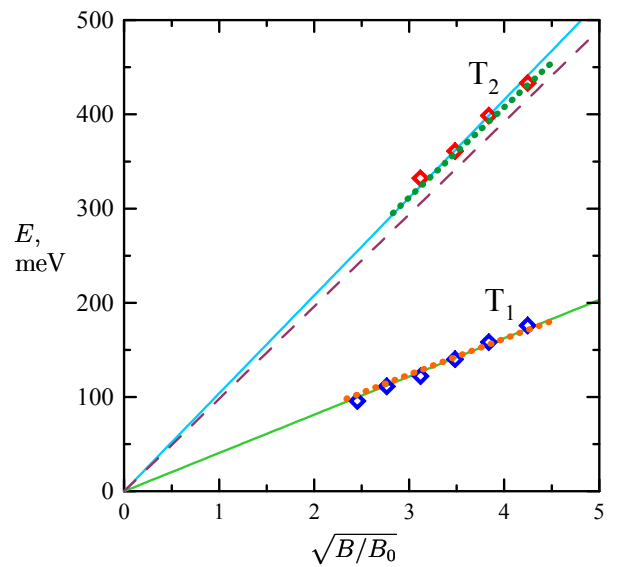}}
\end{center}
\caption{\label{Fig4}Energies of $\mathrm{T}_1$ and $\mathrm{T}_2$ cyclotron transitions as functions of square root of
magnetic field (in the units of $B_0=1\,\mbox{T}$). Diamonds: the experimental data \cite{Jiang1,Jiang2}, solid lines:
the linear fits to the experimental points (see Table~\ref{Table2}, first column), dotted lines: the best theoretical
fit with the self-consistent screening (see Table~\ref{Table2}, last column). Dashed line: the $\mathrm{T}_1$ linear
fit, multiplied by $1+\sqrt2$, which must pass through the $\mathrm{T}_2$ points in the absence of many-body effects.}
\end{figure}

Experimental data \cite{Jiang1,Jiang2} on the energies of cyclotron $\mathrm{T}_1$ and $\mathrm{T}_2$ transitions in
graphene are depicted in Fig.~\ref{Fig4} together with their linear fits $E\propto\sqrt{B}$. In these experiments,
graphene is placed onto a $\mathrm{SiO}_2$ substrate and is electrostatically doped up to a complete filling of the
$n=0$ ($\nu=2$) or $n=-1$ ($\nu=-2$) Landau level. The many-body effects are seen in deviation of the $\mathrm{T}_2$
energies from the $\mathrm{T}_1$ ones multiplied by $1+\sqrt2$ (dashed line in Fig.~\ref{Fig4}).

Our fits of the whole set of experimental points obtained by adjustment of $\varepsilon$ at fixed
$v_\mathrm{F}=0.85\times10^6\,\mbox{m/s}$ in different approximations are described in Table~\ref{Table2}. In the case
of unscreened interaction, we get, as before, overestimated $\varepsilon$ and insufficient interline distance (i.e. the
difference between $v_\mathrm{F}^*$ for $\mathrm{T}_1$ and $\mathrm{T}_2$ is smaller than in the experiment). Taking
into account the screening, we obtain much lower $\varepsilon$ and still insufficient interline distance. Finally,
self-consistent treatment of the screening improves agreement with the experiments, as shown by dotted lines in
Fig.~\ref{Fig4}.

\begin{table*}[t]
\centering
\begin{tabular}{|c||c|c|c|c|}
\hline%
&Experiment&Unscreened&Screened&Self-consistent\\
&\cite{Jiang1,Jiang2}&interaction&interaction&screening\\
&&$r_{\mathrm{s},n}=0$&$r_{\mathrm{s},n}=e^2/\varepsilon v_\mathrm{F}$&$r_{\mathrm{s},n}=e^2/\varepsilon\langle
v_\mathrm{F}^*(\mathrm{T}_n)\rangle$\\
\hline%
$\varepsilon$&---&6.84&2.81&3.86\\
\hline
$v_\mathrm{F}^*(\mathrm{T}_1),\:10^6\,\mbox{m/s}$&1.119&1.136&1.134&1.131\\
\hline
$v_\mathrm{F}^*(\mathrm{T}_2),\:10^6\,\mbox{m/s}$&1.186&1.165&1.168&1.171\\
\hline
\end{tabular}
\caption{\label{Table2}First line: optimal dielectric constants $\varepsilon$ of surrounding media, obtained by least
square fittings of the cyclotron resonance experimental data from Refs.~\cite{Jiang1,Jiang2} at
$v_\mathrm{F}=0.85\times10^6\,\mbox{m/s}$ in different approximations. Second and third lines: the renormalized Fermi
velocities $v_\mathrm{F}^*$, which are extracted from the experimental data and calculated at optimal $\varepsilon$ on
average over the experimental magnetic field range.}
\end{table*}

\section{Conclusions}
\label{sec4}

We present a theoretical study of many-body effects of Coulomb interaction in graphene in strong magnetic field.
Calculating the energies of the experimentally observable inter-Landau level transitions, we consider the
single-particle self-energies, the excitonic effects, and the Landau level mixing. Moreover, this work presents the
first systematic calculation which takes into account the screening of the Coulomb interaction and is aimed on detailed
comparison with experiments.

The analysis of the experimental data \cite{Berciaud,Faugeras} on magneto-Raman $\mathrm{L}_1$ and $\mathrm{L}_2$
transitions in graphene on three different substrates has resulted in the following conclusions:

a) The optimal value of the bare Fermi velocity is about $v_\mathrm{F}=0.85\times10^6\,\mbox{m/s}$, in agreement with
the estimates $(0.8\div0.9)\times10^6\,\mbox{m/s}$ of this quantity based on analysis of recent experimental data
\cite{Yu,Hwang2,Faugeras,Lozovik2}.

b) Calculations with the unscreened Coulomb interaction require too large dielectric constants $\varepsilon$ to obtain
the best least square fits of experimental points and cannot accurately reproduce the distances between $\mathrm{L}_1$
and $\mathrm{L}_2$ spectral lines.

c) Static screening of the interaction yields too low values of $\varepsilon$ and also underestimates the interline
distances.

d) Self-consistent treatment of the screening, which approximately models renormalization of transition energies in
electron polarizability \cite{Note2}, greatly improves agreement of calculations with the experiment. The best fits are
achieved when $\varepsilon\approx2.2$ for suspended graphene, $\varepsilon\approx4.5$ for graphene in hBN, and
$\varepsilon\approx8$ for graphene on graphite.

The similar conclusions are made from analysis of the experiments \cite{Jiang1,Jiang2} on cyclotron $\mathrm{T}_1$ and
$\mathrm{T}_2$ transitions in graphene on a $\mathrm{SiO}_2$ substrate. Our best theoretical fit is achieved with the
self-consistent screening at $v_\mathrm{F}=0.85\times10^6\,\mbox{m/s}$ and $\varepsilon\approx3.9$.

To conclude, we analyze the experimental data on many-body signatures in inter-Landau level transitions in graphene in
strong magnetic fields. We show that taking into account the self-consistent screening of Coulomb interaction plays the
key role in achieving agreement between the theory and experiments.

Our approach will be further developed by considering dynamical effects in the interaction screening in a subsequent
work. It is also interesting to analyze possible signatures of Coulomb many-body effects observed in graphene on SiC
\cite{Miller}, hBN \cite{Chen}, GaAs, and glass \cite{Takehana} substrates in strong magnetic fields.

\section*{Acknowledgements}
The work was performed with financial support of Russian Ministry of Science and Education (Project \#RFMEFI61316X0054).

\end{document}